\documentclass[a4paper, oneside, twocolumn, notitlepage, 10pt]{extarticle_ecoc}
\usepackage{ecoc}

\begin{document}
\selectlanguage{english}    


\title{Silicon Photonics Wavelength-Independent C-Band Tunable Optical Filter With Feasible Thermal Tuning Requirements}%


\author{
    Saif Alnairat\textsuperscript{(1, 2)}, Benjamin Wohlfeil\textsuperscript{(1)}, Stevan Djordjevic\textsuperscript{(1)}, Bernhard Schmauss\textsuperscript{(2)} 
}

\maketitle                  


\begin{strip}
 \begin{author_descr}

   \textsuperscript{(1)} ADVA Optical Networking SE, M{\"a}rzenquelle 1, 98617 Meiningen, Germany,
   \textcolor{blue}{\uline{salnairat@adva.com}} 
   
   \textsuperscript{(2)} LHFT, Friedrich-Alexander Universit\text{\"a}t Erlangen-N\text{\"u}rnberg, 91058 Erlangen, Germany.
  
 \end{author_descr}
\end{strip}

\setstretch{1.1}
\renewcommand\footnotemark{}
\renewcommand\footnoterule{}


\begin{strip}
  \begin{ecoc_abstract}
    A filter design based on Vernier microrings and wideband directional couplers is proposed for ASE noise suppression in next generation DCI applications. We demonstrate a $\sim$40 nm FSR-free filter with $>$ 20.5 dB average ER and 3dB-BW of $\sim$75 GHz, achieving wavelength-independent performance and full tunability with a maximum tuning temperature of $\sim$75 K. 
    \textcopyright2022 The Author(s)
  \end{ecoc_abstract}
\end{strip}


\section{Introduction}
\vspace{-0.5em}
In order to deploy $>$ 25 Tbit/s transmission systems in next-generation Data Center Interconnect (DCI) networks, performance challenges have to be resolved\cite{example:conference0, example:article2}. One of the most critical challenges facing high-speed, filterless, i.e., with no Arrayed Waveguide Grating (AWG) filter, Dense Wavelength-Division Multiplexing (DWDM) transceivers is Optical Signal to Noise Ratio (OSNR) degradation, due to accumulated Amplified Spontaneous Emission (ASE) noise originating from optical amplifiers. Thus, tunable ASE optical filters are essential components for advanced optical transceivers and transmission systems to achieve the OSNR tolerance threshold.

Micro-Ring Resonators (MRRs) have been widely used in communication applications
due to thier small footprint and scalability\cite{example:conference2, example:article1}. Over the past decade, various integrated silicon MRRs-based optical filters have been explored and different types of configurations have been analyzed. Tunable optical filters based on series coupled and cascaded MRRs in addition to Vernier filters with a tuning range of up to 124 nm were fabricated and reported\cite{example:article3, example:article4, example:article5, example:article6, example:article7, example:conference1}. However, designs tunable over $>$ 35 nm range either require very high tuning temperature {$\Delta{T}$} =\: (125 - 400) K, or exhibit insufficient 3-dB bandwidth (3dB-BW), low extension ratio (ER) and high insertion loss (IL) to pass through single wavelength channels while simultaneously suppressing the remaining C-band. Furthermore, the performance of these designs varies across their tuning range due to chromatic dispersion and wavelength dependency of the MRRs couplings strength across the tuning range\cite{example:article4, example:conference1}. In addition to varying performance, under-coupling increases the complexity of resonance alignment and automatic tuning due to resonance splitting\cite{example:article4,example:article5,example:article6}.

In this work, we propose a novel filter design based on Vernier MRRs and wideband directional couplers, to overcome thermal challenges and wavelength-dependent performance. A two-stage coupled MRR Vernier wavelength-independent filter example design suitable for dual-polarization (DP) 64-Quadrature Amplitude Modulation (QAM)/64-GBd systems is presented, and its performance is compared to a conventional Vernier filter design. The filter is automatically tuned achieving (73 - 75) GHz 3dB-BW and 20.5 dB average ER over the C-band with a maximum {$\Delta{T}$} =\: $\sim$75 K, realizing wavelength-independent performance.    
\vspace{-1em}
\begin{figure*}[t]
   \centering
    \includegraphics[width=0.98\linewidth]{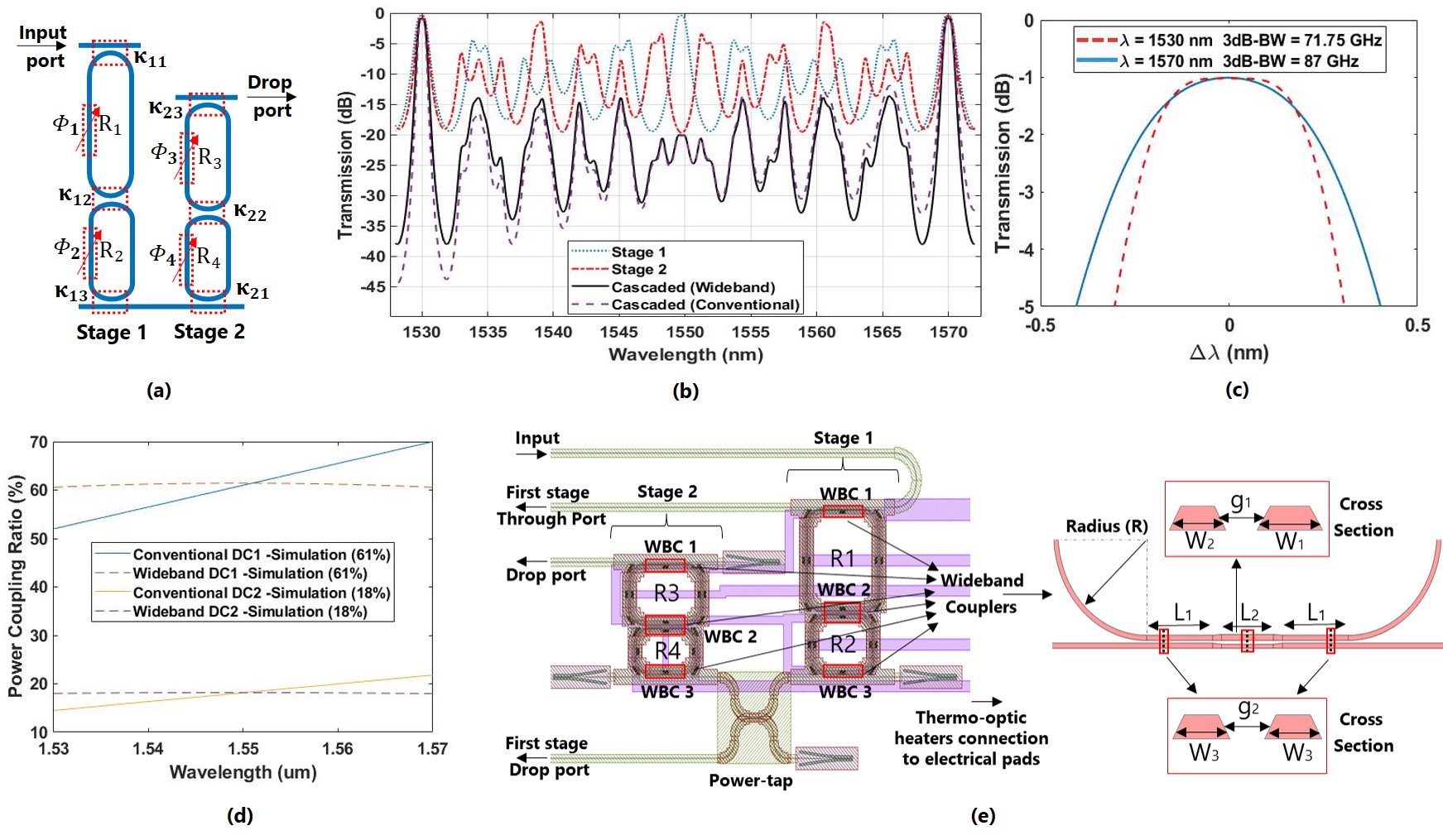}
    \caption{(a) Schematic of a two-stage coupled MRR Vernier filter. Thermal Phase shifters ($\phi_i$) and ring-couplers ($\kappa_i$) for both stages are highlighted. (b) Comparison of the simulated drop port transmission spectra of each stage of the Vernier filter and the spectra of the cascaded filter with conventional and wideband directional couplers. (c) Close-up comparison of the drop port peak at 1.53 and 1.57 $\mu$m for the conventional cascaded filter design. (d) Simulated cross splitting ratio of conventional and wideband directional couplers for the inner and outer ring-couplers. (e) The layout of the fabricated wavelength-independent filter and a schematic of the wideband ring-couplers.}
    \label{fig:figure1}
\end{figure*}
\section{Design optimization}
The required properties of the filter depends on the application and the requirements of the system where it is employed. The design approach was first to model an AWG-less DWDM optical communication system which utilizes DP-64-QAM/64-GBd transmitters to investigate and analyze the required properties of the ASE filter, including performance in the aspect of OSNR requirement of the system. The model shows that for a 66-channel system, an average ER of $>$ 17 dB is required to achieve the OSNR threshold for the mentioned system. Thus, the design targets 20 dB average ER and 80 GHz 3dB-BW to increase spectral misalignment tolerance. The design configuration is chosen to achieve the desired characteristics taking into account design limitations, including tunability, thermal challenges, complexity, power consumption and tolerance to fabrication variations. 

Fig.~\ref{fig:figure1}(a) shows a schematic for the design configuration with first and second stages being composed of two different second-order Vernier filters. Considering a silicon-on-insulator (SOI) platform for this design and assuming thermo-optic coefficient of 1.86 $\times$ 10$^{-4}$ K$^{-1}$ with a maximum {$\Delta{T_{heater}}$} =\: $\sim$75 K, which would give a {$\Delta{T_{wg}}$} =\: $\sim$69 K based on thermal simulation for the designed phase shifters, the minimum resonator-length ($L_{min}$) allowed to achieve 2$\pi$ phase shift is 121 $\mu$m. Giving the group index at 1.55 $\mu$m is 3.96, this gives a free spectral range (FSR) of $\sim5$ nm for $L_{min}$. The lengths of different resonators on both filters are chosen such that the total required FSR, i.e. 40 nm, is the least common multiple of the FSR of individual resonators, and the secondary resonances of the cascaded filter output are mutually suppressed, as shown in Fig.~\ref{fig:figure1}(b). The FSRs of \{$R_{1}$, $R_{2}$, $R_{3}$, $R_{4}$\} are \{$3.08$, $4.44$, $4$, $5$\} nm, corresponding to \{$L_{R1}$, $L_{R2}$, $L_{R3}$ $L_{R4}$\} of \{$197.17$, $136.51$, $151.67$, $121.34$\} $\mu$m, respectively. To simplify the design, we choose (${{\kappa}_{out}} = {{\kappa}_{11}} = {{\kappa}_{13}} = {{\kappa}_{21}} = {{\kappa}_{23}}$) and (${{\kappa}_{in}} = {{\kappa}_{12}} = {{\kappa}_{22}}$). In this case, the coupling strengths required to achieve the targeted 3dB-BW and average ER with a flat bandpass at 1.55 $\mu$m were found to be ${{\kappa}_{in}} = 0.18 $ and ${{\kappa}_{out}} = 0.61$.

Due to chromatic dispersion and wavelength dependency of the power couplings for conventional couplers designs, under and over coupling usually occur, which leads to narrower 3dB-BW with higher ER and larger 3dB-BW with lower ER at the lower and upper end of the spectrum, respectively, as shown in Fig.~\ref{fig:figure1}(b) and Fig.~\ref{fig:figure1}(c). 
\begin{figure*}[t]
   \centering
    \includegraphics[width=0.98\linewidth]{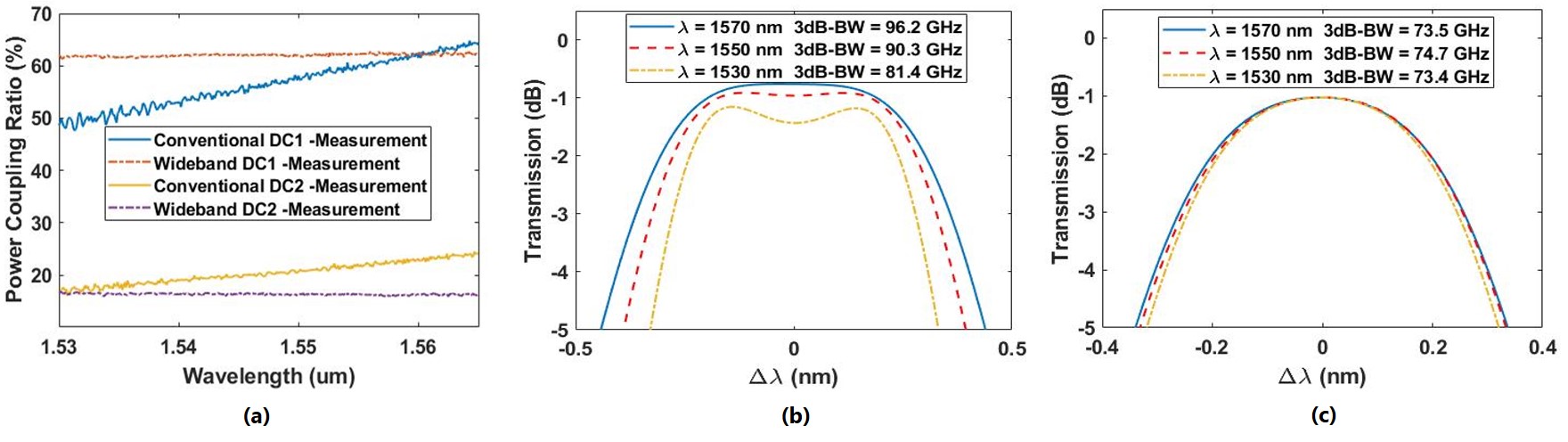}
    \caption{(a) Measured cross splitting ratio of conventional and wideband directional couplers for the inner and outer ring-couplers. (b) Close-up comparison of the measured drop port peak at 1.53, 1.55 and 1.57 $\mu$m for the conventional cascaded filter design. (c) Close-up comparison of the measured drop port peak at 1.53, 1.55 and 1.57 $\mu$m for the wideband cascaded filter design.}
    \label{fig:figure2}
\end{figure*}

To realize uniform coupling strength over the entire tuning range, inner (ring-ring) and outer (bus-ring) wideband couplers are used. Fig.~\ref{fig:figure1}(d) shows simulated cross coupling splitting ratios for the designed conventional and wideband outer and inner ring-couplers, where conventional couplers based on symmetric straight couplers show $\pm$14$\%$ variation across the tuning range. The wideband couplers are designed using asymmetric-waveguide based phase control sections to introduce a small phase shift between the two symmetric couplers and compensate for the wavelength-dependent power coupling \cite{example:article10}. In order to realize inner wideband ring-coupler, two different outer wideband ring-couplers designs are needed. 

Fig.~\ref{fig:figure1}(e) shows the layout of the fabricated wideband filter and the schematic of the wideband coupler. All couplers have similar bend radius $R$ = 10 $\mu$m, taper length = 1 $\mu$m, $W_1$ = 400 nm, $W_2$ = 600 nm and $W_3$ = 500 nm. The rest of the design parameters are summarized in Tab.~\ref{tab:table1}. The resonator lengths of the conventional design are slightly adjusted to take into consideration the change of the effective index in the phase section. To increase tuning flexibility, a power-tap coupler is added at the drop port of the first stage. 

\begin{table}[h]
\centering
\caption{Design parameters of wideband couplers} \label{tab:table1}
\begin{tabular}{|c|c|c|c|}
\hline
\bf{Parameter}  & \bf{WBC1} & \bf{WBC2} & \bf{WBC3}  \\
\hline
$g_1$ ($nm$) &  350&  300&  300\\
\hline
$g_2$ ($nm$)  & 250&  200&  200\\
\hline
$L_1$ ($\mu$$m$) & 12.51&  7.16&  7.58\\
\hline
$L_2$ ($\mu$$m$) & 2.96&  5.8&  5.1\\
\hline
${\kappa}$ (${\%}$) & 61&  18&  61\\
\hline

\end{tabular}
\end{table}

\section{Measurement, Results and Discussion}
To measure the cross splitting ratio of the fabricated conventional and wideband directional couplers, a continuous tunable C-band laser and a power meter were coupled to the couplers test structures using a fiber array and grating couplers. The laser wavelength was swept with 0.1 nm step across (1530 nm - 1565 nm) wavelength range and the power reading was normalized to the grating coupling loss. Fig.~\ref{fig:figure2}(a) shows the cross splitting ratio of conventional and wideband couplers. The inner and outer ring-couplers measured splitting ratios at 1550 nm for conventional and wideband designs are \{0.205, 0.58\} and \{0.163, 0.62\}, respectively, which means that the conventional filter design is under-coupled while the wideband filter design is slightly over-coupled. Conventional couplers show $\sim$ $\pm$15$\%$ coupling variation across the tuning range, while wideband couplers show only $\pm$1$\%$ variation. 

To measure and tune the filter spectra, the input port was connected to the laser, and the drop ports for the first stage, i.e., the power tap, and the cascaded filter were connected to photodetectors and feedback control circuitry. The tunable laser source was set to the desired resonance wavelength and an automatic resonance alignment algorithm was applied to tune the optical filter. A combination of modified derivative-based and derivative-free optimization algorithms were adapted based on thermal characterization of the device. An additional step was required to fine tune the conventional filter spectrum due to resonance splitting. The IL of conventional power-tap coupler was subtracted from the measured transmission. Fig.~\ref{fig:figure2}(b) shows the measured drop port peak at several resonance wavelengths for the conventional filter design. The filter has a $\sim$40 nm FSR, an average ER of 21 dB and a minimum out-of-band suppression of 14 dB. However, the difference in 3dB-BW between the upper and lower end of the C-band is around 15 GHz, while IL varies between 0.9 and 1.4 dB. Furthermore, the resonance splitting due to under-coupling is evident at the lower end of the tuning range. On the other hand, Fig.~\ref{fig:figure2}(c) shows the measured drop port peak for the wideband filter design, where the difference in 3dB-BW across the tuning range is less than $<$ 1.5 GHz. The IL is around 1 dB while the average ER is 20.5 dB. The resonance alignment requires only 250 iterations for each tuning process, which is much faster than previously reported for similar device configurations\cite{example:article4}. The power required to tune each thermal phase shifter by 2$\pi$ is around 80 mW. To increase tolerance to fabrication variations, outer wideband couplers could be designed such that they have similar phase section but different bend radii as well, while keeping the current design for the inner ring-coupler. 
\section{Conclusions}
 In this paper, we have proposed and experimentally demonstrated a two-stage second-order MRR Vernier wavelength-independent filter based on wideband directional couplers. The device was automatically tuned across the C-band, achieving $<$ 1.5 GHz 3-dB bandwidth variation and  realizing wavelength-independent performance with feasible thermal tuning requirements.

\section{Acknowledgements}
This work has been partially funded by the German Ministry of Education and Research under the grant agreement 13N14937 (PEARLS).


\printbibliography

\vspace{-4mm}

\end{document}